# Chapter 5. Analytical Models of Bulk and Quantum Well Solar Cells and Relevance of the Radiative Limit

**James P. Connolly**

*Nanophotonic Technology Centre, Universidad Politécnica de Valencia, Camino de Vera s/n, 46022 Valencia , Spain*

**ABSTRACT**

The analytical modelling of bulk and quantum well solar cells is reviewed. The analytical approach allows explicit estimates of dominant generation and recombination mechanisms at work in charge neutral and space charge layers of the cells. Consistency of the analysis of cell characteristics in the light and in the dark leaves a single free parameter, which is the mean Shockley-Read-Hall lifetime. Bulk PIN cells are shown to be inherently dominated by non-radiative recombination as a result of the doping related non-radiative fraction of the Shockley injection currents. Quantum well PIN solar cells on the other hand are shown to operate in the radiative limit as a result of the dominance of radiative recombination in the space charge region. These features are exploited using light trapping techniques leading to photon recycling and reduced radiative recombination. The conclusion is that the mirror backed quantum well solar cell device features open circuit voltages determined mainly by the higher bandgap neutral layers, with an absorption threshold determined by the lower gap quantum well superlattice.

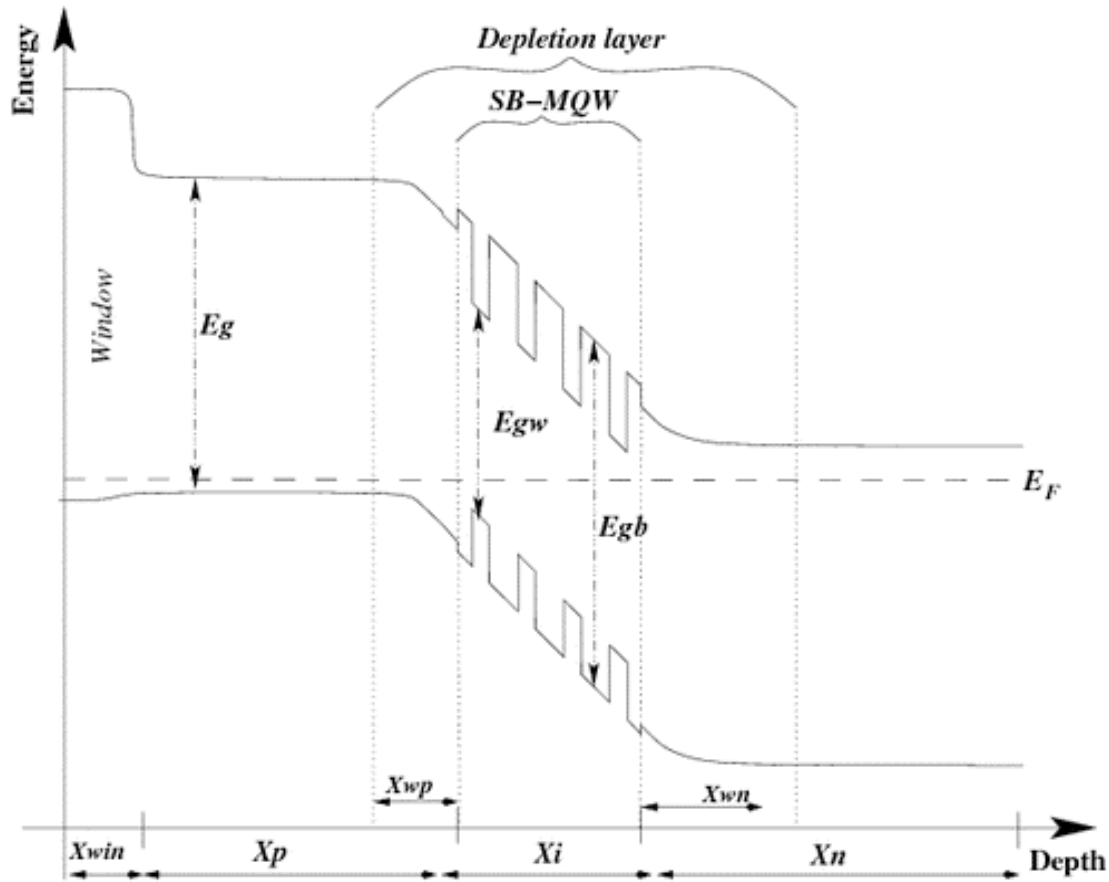

*Figure 5.1. The strain balanced quantum well solar cell (SB-QWSC) structure. Alternating strain balanced wells and barriers of gaps Egw, Egb make up the intrinsic region of total width Xi, which is sandwiched between p and n doped layers of width Xp, Xn and bandgap Eg, with an optional higher bandgap window layer. Widths are not to scale, and typical QWSCs contain some tens of QW periods.*

## 1. Introduction

Despite great advances in physical understanding, in materials, and in fabrication, and despite reaching efficiencies over 40%, just two routes to higher efficiencies have been comprehensively studied. The first is the multi-junction cell concept, which reduces thermalisation losses. The second is the use of light concentration, reducing the solid angle for light emission towards the minimum which is the angle for light acceptance (A. de Vos, 1992). The current interest in novel phenomena, often involving nanostructures, is part of the effort to go beyond these early ideas.

This chapter investigates generation and loss mechanisms in bulk and quantum well solar cells, with emphasis on developing physical understanding via analytical models, rather than more accurate but less revealing numerical methods. The designs studied are bulk PIN cells contrasted with quantum well solar cells (QWSCs), complementing other nanostructured designs reviewed in other chapters.

The quantum well solar cell (QWSC, figure 5.1) is a *p-i-n* or *n-i-p* solar cell design with quantum wells (QWs) in the undoped intrinsic *i* region (K.W.J. Barnham and G. Duggan, 1990). Carrier escape studies show efficient field assisted thermal escape of the order of picoseconds, for carrier lifetimes of several nanoseconds (Nelson, J.; Paxman, M.; Barnham, K.W.J.; Roberts, J.S.; Button, 1993). Early quantum efficiency (QE) modelling (Paxman *et al.* 1993) further shows that escape efficiency is essentially 100%. More recent theoretical work by Chin-Yi Tsai and Chin Yao Tsai (2007) confirms this while showing that escape times must be at least two orders of magnitude shorter than recombination lifetimes for a net efficiency again to be achievable.

Consequently, however, it is clear that this efficient collection requires that the field be maintained across the wells. The nominally undoped wells and barriers however inevitably contain a net background doping level, corresponding to a fixed charge density which increasingly degrades the built in field, the wider the multiple quantum well (MQW) superlattice is grown. This brings us to the first design issue with these cells, which is the practical upper limit on total intrinsic region thickness *Xi* and corresponding limit on absorbing MQW thickness that may be fabricated. This materials quality dependent limit may extend well over 1μm, and even in direct gap quantum wells makes this system well suited to light trapping techniques as we will see in subsequent sections.

More recent analytical models by Rimada, Hernàndez, Connolly and Barnham (2007) have followed a similar analytical methodology confirming early results that an MQW can enhance efficiency in non-ideal, high bandgap cells, but do not demonstrate an advantage for ideal material. The MQW bandgap together with near unit collection efficiency leads to a net increase in short circuit current (Isc). This increase in Isc however is accompanied by an increase in recombination in the low gap well regions, as discussed in some detail by Anderson (1995) for example.

Bearing this in mind, the first non trivial advantage of the QWSC is one of materials. The alternation of thin barrier and well regions allows the use of alternating tensile and compressive materials: this is the strain balancing technique (Barnham *et al.* 2006) allowing a much wider range of materials to be explored without dislocations. This variation is the strain balanced SB-QWSC, illustrated in figure 5.1 with the strain balanced quantum well superlattice or SB-MQW.

The second potential advantage is the result of the materials inhomogeneity and consequent departure from the homogeneous pn junction picture of solar cells. The difference in carrier properties and carrier dynamics between wells and barriers have lead to much discussion concerning the quasi-equilibrium concentration of carriers in the wells and barriers. In particular, the possibility of suppressed quasi-Fermi levels in the wells and higher carrier temperatures has been investigated by studies of steady state luminescence, as reviewed by Barnham *et al.* (2006) and developed by Connolly *et al.* (2007).

Finally, the geometry of the QWSC introduces a further inhomogeneity that demonstrably leads to efficiency enhancement, which is the inhomogeneous luminescence from bulk and quantum well layers, and it's potential exploitation for reduced recombination losses in practical devices.

In order to investigate this, the following sections review an analytical QWSC model of QWSC allowing estimation of different loss mechanisms in a QWSC, and comparing and contrasting these with the case of bulk solar cells.

## 2. Analytical model

The modelling methodology uses analytical solutions to allow explicit description of physical phenomena. The use of analytical methods rather than more exact numerical methods allows development of greater physical understanding. This comes at the expense, however, of accuracy and generality as a result of approximations required. This method is chosen since the prime focus is developping understanding rather than optimising devices, for which numerical methods are preferred. The modelling we develop nevertheless focusses on quantitative modelling of experimental data, and measured data is preferred to ab initio calculation of parameters wherever possible.

In this section we will describe the components of the model SOL starting with notes on the description of structural and physical cell parameters which includes quantum well density of states and absorption coefficient calculations. These notes lay the foundation of the QE calculation first reported by Paxman *et al.* (1993) and followed by the integration of non-radiative (Connolly *et al.* 2000) and radiative (Connolly *et al.* 2004) mechanisms and finally light trapping and photon recycling elements (Connolly *et al.* 2007).

The method makes use of the complementarity of generation and recombination phenomena. This complementarity typically fixes all sample parameters except the non radiative lifetime. For example the QE determines cell absorption and charge-neutral layer minority carrier lifetimes, which in turn quantitatively determine radiative and non radiative recombination currents.

Concerning the structures, we make the common assumptions of homogeneous composition in doped and intrinsic layers, the depletion approximation in the space-charge region (SCR), and 100% photogenerated carrier collection in the SCR.

The first cell characteristic simulated is the spectral response (SR), yielding the cell QE and short circuit current (Isc) for a given spectrum. The fit to the QE determines the recombination characteristics independently in charge neutral and space-charge regions. This determines the radiative and non radiative recombination currents in these regions as a function of applied bias.

The overall photocurrent is simply expressed in terms of superposition, adding photocurrent to the dark current in order to ascertain the light current characteristic.

## 2.1 Bulk and quantum well model

The materials of interest are the $Al_{(x)}Ga_{(1-x)}As$ and $In_{(x)}Ga_{(1-x)}As_{(y)}P_{(1-y)}$ families and their binary, ternary and quaternary compounds. For the modelling reviewed here, the materials parameters of there materials in the bulk rely exclusively on the rich literature in this field (Vurgaftman & Meyer 2001, EMIS datareviews series 1990 no. 2, EMIS datareviews series 1991 no. 6, S. Adachi 1994, S. Adachi 1992, O. Madelung 1996).

The quantum well parameters are calculated in the finite square well picture under the effective mass approximation as described in detail by Nelson (Jenny Nelson *et al.*, 1999) and references therein. We summarise and extend the method here in order to define assumptions and relevant parameters.

The wells we are interested in are significantly greater than the lattice periodicity allowing us to use the envelope function approximation. We further assume that the wavefunction in the un-confined plane of the well and in the growth direction are decoupled, such that the confinement may be assumed one-dimensional and the effective mass may be assumed equal to the growth-direction value. The problem is then that of the one-dimensional solution of the Schrödinger equation for a finite square well.

The well depth is evaluated from the bandgap and band offsets of barrier and well materials in the bulk, and the masses of the three carriers estimated where possible from the literature and calculated from k.p methods otherwise (M.C. Lynch *et al.* 2006). In the envelope function approximation, the effective Shrödinger equation takes the form

$$\left[-\frac{h^2}{8\pi m^*}\nabla^2 + V(x)\right]\varphi(x) = E_n\varphi(x) \tag{5.1}$$

for the envelope function φ, carrier effective mass $m^*$, potential $V$ and energy level $E_n$ at position $x$. The boundary conditions, after G. Bastard (1988), are continuity of well and barrier wavefunctions at the well-barrier interface $x_i$, and continuity of the gradient of the wavefunctions at the same interface

$$\varphi_W(x_i) = \varphi_B(x_i) \qquad \frac{1}{m_W^*}\frac{\partial}{\partial x}\varphi_W(x_i) = \frac{1}{m_B^*}\frac{\partial}{\partial x}\varphi_B(x_i) \tag{5.2}$$

Solving of the Schrödinger equation equation (1) subject to these boundary conditions (5.2) gives even and odd solutions

$$\frac{k}{m_W^*}\tan(kL/2) = -\frac{c}{m_W^*} \qquad \frac{k}{m_W^*}\cot(kL/2) = -\frac{c}{m_W^*} \tag{5.3}$$

in terms of wavevector k, well width L and extinction coefficient c which is found by solving equations 5.3 numerically. This yields bound electron and hole wavefunctions in the well described by potential profile V(x). The resulting absorption coefficient is found from Fermi's Golden Rule defining the probability of transitions from initial to final bound states with an additional excitonic factor $f_i$ accounting for electron-hole Coulomb interaction

$$|f_i|^2 = \frac{R_{CV}}{(1-\nu)^3} \tag{5.4}$$

in terms of the effective Rydberg for the material $R_{CV}$ and $\nu$ the bi-dimensionality of the exciton. Finally, the absorption for wells of index $n$ at angular frequency ω is

$$\alpha(\omega) = \frac{4\pi e^2}{ncm^2\omega}\sum_i\sum_f\left|<\varphi_{v,i}|\varphi_{c,f}>\right|^2\left(\frac{\pi\mu_{i,f}^*}{h^2}\right)|e.M_{CV}|^2$$

$$\times\sum_i\left(|f_i|^2\delta(\frac{h\omega}{2\pi} - E_i) + \Theta\left(\frac{h\omega}{2\pi} - E_g - E_i\right)\right) \tag{5.5}$$

where $\mu_{CV}^*$ is the reduced effective mass for initial $i$ to final $f$ states, $m$ the free electron mass, and $M_{CV}$ is the bulk conduction to valence band matrix element, defined for light and heavy hole transitions by

$$|eM_{CV}^{hh}|^2 = 3|eM_{CV}^{lh}|^2 = \frac{E_P}{4m} \tag{5.6}$$

$E_p$ in eV expresses the interband Kane matrix element, and is a known tabulated materials parameter (Vurgaftman 2001).

The quantum well modelling introduces two free fitting parameters which are the excitonic strength and broadening $\nu$ and the absolute well absorption strength scaling all transitions and accounting for uncertainties in effective densities of states and systematic uncertainties in the model.

## 2.2 Photocurrent

The methodology yielding the photocurrent follows standard methods described by Hovel (1975) for example. It differentiates between photogenerated carrier collection mechanisms in diffusion dominated charge neutral layers and drift dominated space-charge layers.

### 2.2.1 Charge neutral photocurrent

In the charge neutral layers, the spectral response (SR) is calculated as a function of wavelength by solving drift dominated transport and continuity equations. Cell dimensions are defined in Fig. 5.1. The doping densities and known intrinsic carrier densities of the homogeneous $p$ and $n$ layers define the depletion widths $x_{wp}$ and $x_{wn}$ by analytically solving Poisson's equation in terms of fixed charge density only in the depletion approximation.

The SR calculation method used is standard (Hovel 1975) and is briefly summarised here for completeness for a $p$-$i$-$n$ structure. The generation at position $x$ and photon wavelength $\lambda$ of incident light flux $F$ and front surface reflectivity $R$ is

$$G(x,\lambda) = F(1-R)e^{-\alpha x} \tag{5.7}$$

in homogeneous material with absorption coefficient $\alpha(\lambda)$. In the absence of an electric field in the charge neutral layer, current transport is diffusive only. The generalisation to a structure with a planar back surface mirror as discussed below is a straightforward in terms of an infinite geometric sum of optical paths determining the total generation rate.

Current continuity defines the minority carrier densities $n_p$ and $p_n$ in $p$ and $n$ layers respectively as follows

$$\frac{d^2 n_p}{dx^2} = \frac{n_p}{L_n^2} - \frac{G(x)}{D_n} \tag{5.8a}$$

$$\frac{d^2 p_n}{dx^2} = \frac{p_n}{L_p^2} - \frac{G(x)}{D_p} \tag{5.8b}$$

where $L_n$ and $L_p$ minority carrier diffusion lengths and $D_n$ and, $D_p$ minority carrier diffusion constants in $p$ and $n$ layers respectively. Eqns. 5.8a and 5.8b are solved analytically for the p layer subject to minority carrier current density Jn at front surface position $x_w$ (see fig. 5.1) determined by a surface recombination velocity $Sn$, and the depletion approximation at the SCR of vanishing minority carrier concentration due to the built-in junction potential

$$j_n(x) = q S_n n_p(x) \quad (\text{x=x}_{\text{win}}) \tag{5.9a}$$

$$n_p(x) = 0 \quad (\text{x= x}_{\text{win}}+\text{x}_{\text{p}}-\text{x}_{\text{wp}}) \tag{5.9b}$$

Similarly for the n layer the boundary conditions at the SCR and at the back surface in terms of minority hole current density $J_p$ and recombination velocity $S_p$ are

$$j_p(x) = q S_p p_n(x) \quad (\text{x=x}_{\text{w}}+\text{x}_{\text{p}}+\text{x}_{\text{i}}+\text{x}_{\text{wn}}) \tag{5.10a}$$

$$p_n(x) = 0 \quad (\text{x=x}_{\text{w}}+\text{x}_{\text{p}}+\text{x}_{\text{i}}+\text{x}_{\text{n}}) \tag{5.10b}$$

Eqns. 5.8a,b, 5.9a,b and 5.10a,b provide analytical solutions for the electron and hole minority carrier profiles in *p* and *n* layers. The SR is then given by the minority carrier density gradient at the *p* and *n* depletion edges respectively. The QE is defined by the SR as the number of charge carriers collected as a fraction of the incident photon flux at a given wavelength.

Generalising to more than a single layer neutral regions follows the same solution methods. The solution piecewise across all homogeneous layers is found using the same boundary conditions (5.9a,b), (5.10,a,b) with additional boundary conditions at each interface between charge neutral layers of continuous charge density and current continuity. The model here uses a single homogeneous base layer and a dual layer emitter in order to include minor short wavelength contributions from the window layer.

In terms of the methodology, we have set out the solutions in order to explicitly define the variables that are set by the SR calculation and data fitting. The surface recombination velocities and transport parameters have similar effects on the SR, but these effects are nevertheless distinguishable if the corresponding losses are significant. That is, a high surface recombination velocity tends to reduce short wavelength response, whereas a short diffusion length reduces photocurrent more evenly over the entire wavelength range. As such, the QE fitting can reliably determine both high surface recombination velocities and low diffusion lengths particularly in the case of cells with poor performance.

Similarly, both parameters become less distinguishable in the case of good quality cells. However, this lesser accuracy applies to cases where the transport losses are negligible. It therefore has little effect on the QE, and on further calculations of recombination currents as we shall see further in the discussion.

## 2.2.2 Space charge region photocurrent

The SR in the space charge region is calculated in the depletion approximation assuming infinite mobility and drift dominated transport, following work by Paxman *et al*. (1993) showing that carriers photogenerated in the space charge region (*SCR*) are collected with close to 100% efficiency as long as the background doping in the i-region is low enough that the built-in field is maintained across the i-region at the operating point. The calculation of the QE of bulk, barrier and QWs therefore reduces to integrating the generation rate across the SCR [3]. The barrier absorption coefficient is again extrapolated from available data in the literature and shifted in energy according to strain, if present (Barnham *et al.* 2006).

The QW absorption is calculated as described above using effective masses for electrons, and light and heavy holes estimated from k.p calculations for strained material, and values from the literature for unstrained material (M.C. Lynch *et al.* 2006). The excitonic strength and broadening, which are growth-dependent parameters, are variables to be fitted to the SR in the well but which which nonetheless remain relatively constant in good material.

## 2.3 Dark current.

## 2.3.1 Charge neutral layer Shockley injection

The QE modelling, as we hve seen, determines values of the minority carrier transport and surface recombination. We can therefore define the Shockley injection current density $J_S$ over the junction in terms of these parameters (Nell and Barnett, 1987) as

$$J_S(V) = q\left(e^{\frac{qV}{K_BT}} - 1\right)\left[\begin{array}{c}\dfrac{n_{ip}^2}{N_A}\dfrac{D_n}{L_n}\left(\dfrac{\dfrac{S_nL_n}{D_n}\cosh\dfrac{x_p}{L_n} + \sinh\dfrac{x_p}{L_n}}{\dfrac{S_nL_n}{D_n}\sinh\dfrac{x_p}{L_n} + \cosh\dfrac{x_p}{L_n}}\right) \\ + \dfrac{n_{in}^2}{N_D}\dfrac{D_p}{L_p}\left(\dfrac{\dfrac{S_pL_p}{D_p}\cosh\dfrac{x_n}{L_p} + \sinh\dfrac{x_n}{L_p}}{\dfrac{S_pL_p}{D_p}\sinh\dfrac{x_n}{L_p} + \cosh\dfrac{x_n}{L_p}}\right)\end{array}\right] \quad (5.11)$$

where $n_{ip}$, $n_{in}$ are the *p* and *n* intrinsic carrier concentrations, $N_A$, $N_D$ the *p* and *n* doping concentrations, $D_n$, $D_p$ minority carrier diffusion constants, and other terms have their usual meaning. Equation 11 includes surface recombination expressed as the diffusion of injected carriers towards the surface of *p* and *n* charge neutral layers.

### 2.3.2 Space charge region non radiative recombination

Calculation of non-radiative recombination via defects in the space charge region follows work developed by Shockley, Read and Hall (Shockley and Read 1952, and R.N. Hall 1952), adapted to QWSC structures by Nelson *et al.* (1999) and further developed as a function of bias for strained materials (Connolly *et al.* 2000, and Lynch *et al.* 2006). The formalism describes the non-radiative recombination rate *U* via mid-gap trapping centres at position *x* in the space charge region as

$$U_{SRH} = \frac{pn - n_i^2}{\tau_n(p + p_t) + \tau_p(n + n_t)} \quad (5.12)$$

where $\tau_n$, $\tau_p$ are electron and hole capture lifetimes, and trap densities $p_t$ and $n_t$ are calculated for the dominant mid-gap traps. Carrier densities *n(x)* and *p(x)* vary according to local potential and local densities of states at position x, across the space charge region (Connolly *et al.* 2000). The calculation is therefore an integral across depleted sections of the p-doped and n-doped layers, barrier, and quantum wells. The non-radiative depletion layer dark current density $J_{SRH}$ is then the integral of *U* over the regions for which the electron and hole Fermi levels separate, namely intrinsic layer width $x_i$ and *p* and *n* depletion widths $x_{wp}$ and $x_{wn}$.

$$J_{SRH} = \int_{x_p - x_{wp}}^{x_p + x_i + x_{wn}} U_{SRH}\,dx \quad (5.13)$$

Hole and electron capture lifetimes are assumed equal in QW and barrier regions and in the absence of deviation from expected dark-current idealities as we shall see in the modelling section. All other parameters are as definedby, and consistent with, the QE calculation.

This method therefore introduces a single free fitting parameter which is the non radiative carrier capture lifetime in the space-charge region, the others being constrained as we have seen.

## 2.3.3 Radiative recombination

The generalised Planck equation expresses light emitted by a grey-body as a function of absorption, geometry, and chemical potential or quasi-Fermi level separation of recombining species. Nelson *et al.* (1999) expressed the electroluminescence of QWSC devices of cylindrical geometry and Connolly *et al.* (2007) extended the formalism in terms to include light trapping and photon recycling.

The generalised Planck equation defines the total luminescent flux from a radiative emitter as an integral over the photon energy as :

$$J_{RAD} = \int_0^\infty \left( q \frac{2n^2}{h^3 c^2} \left( \frac{E^2}{e^{(E-q\Delta\phi)/kT} - 1} \right) \int_S a(E,\theta,s) dS \right) dE \tag{5.14}$$

where n is the refractive index of the grey body, $\Delta\phi$ is the quasi-Fermi level separation, and the other symbols have their usual meanings. The radiative current is the integral of this flux over all energies. The absorptivity $a(E,\theta,s)$ is the line integral along the optical path of radiation at angle *q* with the normal exiting or entering surface *S*.

In the two-dimensional case of the QWSC with front and back surfaces allowing light emission, three paths are possible at both interfaces. A beam striking a surface may be totally internally reflected at angles of incidents greater than the critical angle giving a first internal reflection. For lesser angles, it may be partially reflected, giving a second internal beam, or transmitted.
Tracking these possible absorption and emission pathways allows us to express the total internal sum of light beams analytically as a geometrical sum, where we neglect coherent beams and interference effects, which is a good assumption for the device dimensions (including substrates) considered. The corresponding integral of the absorptivity *a* defines the total absorptivity over the emitting volume $\alpha_s$ which is therefore given as the combination of internal and external geometric sums over all angles (Connolly *et al.* 2007) and takes the form

$$\begin{aligned}\alpha_S &= \int_S a(E,\theta) dS \\ &= 2\pi A_F \left( \int_{\cos\theta_C}^1 \frac{(1-e^{\alpha d \sec(\theta)})(R_F-1)(e^{\alpha d \sec(\theta)}+R_B)}{e^{2\alpha d \sec(\theta)} - R_B R_F} \cos(\theta) d(\cos\theta) \right) \\ &+ 2\pi A_B \left( \int_{\cos\theta_C}^1 \frac{(1-e^{\alpha d \sec(\theta)})(R_B-1)(e^{\alpha d \sec(\theta)}+R_F)}{e^{2\alpha d \sec(\theta)} - R_B R_F} \cos(\theta) d(\cos\theta) \right) \\ &+ 2\pi A_B \left( \int_0^{\cos\theta_C} \frac{e^{\alpha d \sec(\theta)}(R_B-1)(e^{\alpha d \sec(\theta)} - e^{-\alpha d \sec(\theta)})}{R_B - e^{2\alpha d \sec(\theta)}} \cos(\theta) d(\cos\theta) \right)\end{aligned} \tag{5.15}$$

where *d* is the total material thickness, $R_f$ and $R_b$ the front and back surface reflectivities for energy *E*, and $A_F$ and $A_B$ the front and back surface areas. We note the three terms corresponding to total internal reflection, partial internal reflection, and emission through the back surface.

In the Planck grey body formalism, the different contributions to the luminescence inside the emitting volume contribute separately to the overall luminescence. Therefore, we define well, barrier, and charge neutral net absorptivities The Planck formalism describes the emitting grey body as a point source allows us to separate the contributions from different regions in the cell confined between the front and back surface. We can therefore define barrier, well, and charge-neutral absorbances $\alpha_{s_{Well}}$, $\alpha_{s_{Bulk}}$ and $\alpha_{s_{CN}}$ by analogy with equation 15. well and recombination current density corresponding to the net radiative emission from the combined grey body can be expressed as

$$J_{Rad} = \int_0^\infty \left[ \begin{array}{l} \left( \dfrac{\alpha_{s_{CN}}}{e^{(E-q\Delta\phi_{CN})/kT_{CN}} - 1} \right) + \\ \left( \dfrac{\alpha_{s_{Bulk}}}{e^{(E-q\Delta\phi_{Bulk})/kT_{Bulk}} - 1} \right) + \\ \left( \dfrac{\alpha_{s_{Well}}}{e^{(E-q\Delta\phi_{Well})/kT_{Well}} - 1} \right) \end{array} \right] dE \quad (5.16)$$

$$= J_{Rad}^{CN} + J_{Rad}^{Bulk} + J_{Rad}^{Well}$$

in terms of quantum well and bulk quasi-Fermi level separations $\Delta\phi_{CN}$ $\Delta\phi_{Bulk}$ and $\Delta\phi_{Well}$, and carrier temperatures $T_{CN}$, $T_{Bulk}$ and $T_{Well}$.

The terms $J_{Rad}^{CN}$, $J_{Rad}^{Bulk}$ and $J_{Rad}^{Well}$ enable us to explicitly estimate the radiative recombination current from bulk, well as a function of cell geometry and absorption coefficients, subject to knowledge of the quasi-Fermi level separations and temperatures.

In this work, we will assume an equal carrier temperature in all regions. We also assume a constant quasi-Fermi level in well and barrier material, equal to the applied bias. The situation in the charge-neutral layers is more delicate, depending on position and on illumination. An exact solution could be obtained by numerical solution of coupled transport and Poisson equations. In the spirit of analytical solutions considered in this chapter, however, we fall back on another limiting case, which is the case of high mobility in thin charge-neutral layers, or equivalently, diffusion lengths greater than the charge neutral width. In this case, the injected minority carrier concentration remains close to it's value at the depletion edge throughout the charge-neutral layer, and the resulting quasi-fermi level separation remains approximately constant across the charge-neutral width as assumed by Araujo and Martí (1994) for example. In addition to compatibility with these other authors in the field, this assumption is made because it represents the maximum value that the radiative recombination can attain in the charge neutral limit, and therefore sets an upper limit, or best case, for the radiative efficiency of the structures considered.

The Shockley injection we have mentioned (equation 11) combines radiative and non radiative mechanisms in charge neutral layers. We can now combine the upper radiative limit with the Shockley injection in order to explicitly obtain the lowest possible non-radiative recombination rate in the charge-neutral layers which we define as

$$J_S^{CN}(V) = J_S - J_{Rad}^{cn} \quad (5.17)$$

This upper limit for the non-radiative recombination in the charge-neutral layers, combined with the non-radiative SRH rate in the space-charge region gives us the lower limit of non radiative recombination in the solar cell, including surface recombination. We can therefore define a bias dependent radiative efficiency as the ratio of radiative to total dark current as follows

$$\eta_{RAD}(V) = \frac{J_{Rad}}{J_{SRH} + J_{Rad} + J_S^{cn}} \quad (5.18)$$

It is worth emphasising that this is an upper limit on the radiative efficiency of devices, and that devices closest to achieving this limit will be those devices with short diffusion lengths compared with relevant charge neutral widths. We will see in subsequent sections that the most interesting consequences of this analysis occur for the more efficient devices with good minority carrier transport.

## 2.4 Light current and efficiency

The sum of contributions from charge neutral *p*, and *n* zones, and space charge regions gives the total photocurrent density $J_{PH}$. This defines the external quantum efficiency including reflection loss (QE) as ratio of collected carriers to number of incident photons at a given wavelength, that is, the probability that a photon incident on the solar cell gives rise to a charge carrier collected at the cell terminal.

Finally, the light current density under applied bias, assuming superposition of light and dark currents is given by

$$J_L(V) = J_{PH} - (J_S + J_{SRH} + J_{RAD}) \tag{5.19}$$

where we use the photovoltaic sign convention of positive photocurrent.

This light IV enables us in the standard manner (see for example Nelson 2003) to evaluate solar cell figures of merit such as the short circuit current $J_{SC} = J_L(0)$, the maximum power point $V_{mp}$, fill factor FF. Effects of parasitic resistance are included when modelling real data in the usual manner, that is, a series resistance defining a junction bias, and a parallel resistance and associated shunt current reducing the photocurrent.

## 2.5 Discussion

Having sketched the modelling methodology and the separate analytical solutions that, together, describe the overall solar cell performance, the links between the physical phenomena described merit some attention. Firstly, the QE modelling determines all cell parameters except space-charge SRH lifetime. That is, it sets the band structure, absorption profiles, majority carrier densities, and minority carrier transport properties.

The band structure and resulting carrier concentration profiles, together with minority carrier transport determine the Shockley injection current (5.11) across the built in potential.

The carrier densities injected across the SCR in turn determine the SRH non-radiative recombination current (5.13) with the introduction of a free parameter, which is the non radiative capture lifetime introduced in section 5.2.3.

The majority carrier densities and minority carrier transport determine Shockley injection (5.11) as a function of applied bias by diffusion across the built in potential. The same carrier concentration profiles as a function of position across the SCR determine the non radiative SRH recombination profiles (5.13) with the addition of the non-radiative lifetime as single free parameter, which is a sensitive function of growth conditions. This approach has been found to fit data well for biases up to the open circuit voltage (Connolly *et al.* 2007).

These drift-diffusion profiles are consistent with quasi-Fermi level separations determining the luminescence which again are determined by the applied bias.

The Shockley Injection current includes both radiative and non-radiative contributions since both are subsumed into a single minority carrier lifetime or diffusion length. Therefore, the SRH calculation applies only to the SCR region, as does the radiative recombination current. For luminescence calculations, however, the radiative calculation applies to the entire structure in those cases where the diffusion length is comparable or greater than the relevant charge-neutral layer thickness. This is always the case in the high purity material considered, where the dopant species are the only significant defects. A noteworthy feature of this approach is that modelling the QE of the samples uniquely determines all recombination mechanisms except the non-radiative recombination in the SCR. This being determined by a single lifetime, the model therefore fixes excitonic characteristics and minority carrier transport that might otherwise be free parameters. The electroluminescence spectra and hence the radiative contribution to the dark-current are also reproduced essentially without adjustable parameters.

The conclusion we come to after this overview of the interdependence of the analytical solutions is that this analytical approach approach describes recombination mechanisms in detail with no free parameters in the limit of dominant radiative recombination. The reason for this is the linking physical processes common to photogeneration and recombination mechanisms. This explicit description made possible with analytical solutions methods comes at some cost in accuracy resulting from the approximations required. The following analyses of experimental data investigate show whether the understanding gained is justified.

## 3. Radiative limits of bulk versus QWSC

We now investigate bulk PIN and QWSC cases to examine the operating limits of these two structures. Both PIN and SB-QWSC structures are fabricated by MOVPE. The dimensions are similar but for the SB-MQW layer and can be summarised as follows: the PIN consists of a 43nm $Al_{0.7}Ga_{0.3}As$ cap on a 0.5μm GaAs p-type emitter, a 0.9μm intrinsic GaAs layer and finally a 2μm GaAs n-type base layer. The MQW sample consists of a slightly higher bandgap $Al_{0.8}Ga_{0.2}As$ cap, also of width 43nm. This is grown on a slightly thinner 0.4μm GaAs p-type emitter, a 1.46μm intrinsic GaAs layer and finally a 2μm GaAs n-type base layer. The SB-MQW system is made up of $In_{0.11}Ga_{0.89}As$ wells of width 95A, compressively strained between $InGa_{0.911}P_{.089}$ tensile strained barriers of width 196A.

These structures are not optimal structures for conversion efficiency, being instead designed for minimal parasitic resistance effects together with a intrinsic region width ensuring efficient carrier collection at bias up to high bias approaching flat band. As a result, they both comprise a high metallisation and shading fraction, which are 70% and 40% for the PIN and QWSC solar cell devices respectively. In addition, the number of wells in the QWSC sample is not the highest that has been manufactured, leading to a non optimum number of wells. The following discussion therefore will not address specific performance issues, focussing instead on the limiting behaviour we have discussed in section 2.

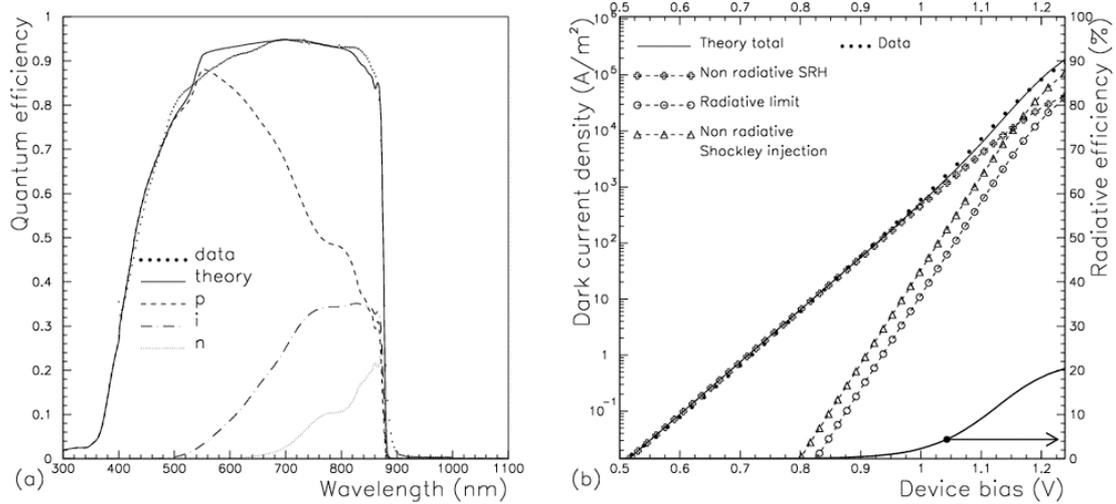

*Figure 5.2: QE data and model for (a) the bulk GaAs PIN and (b) dark current fit together with radiative efficiency peaking at 21% at flat band, and active area efficiency 21.0%.*

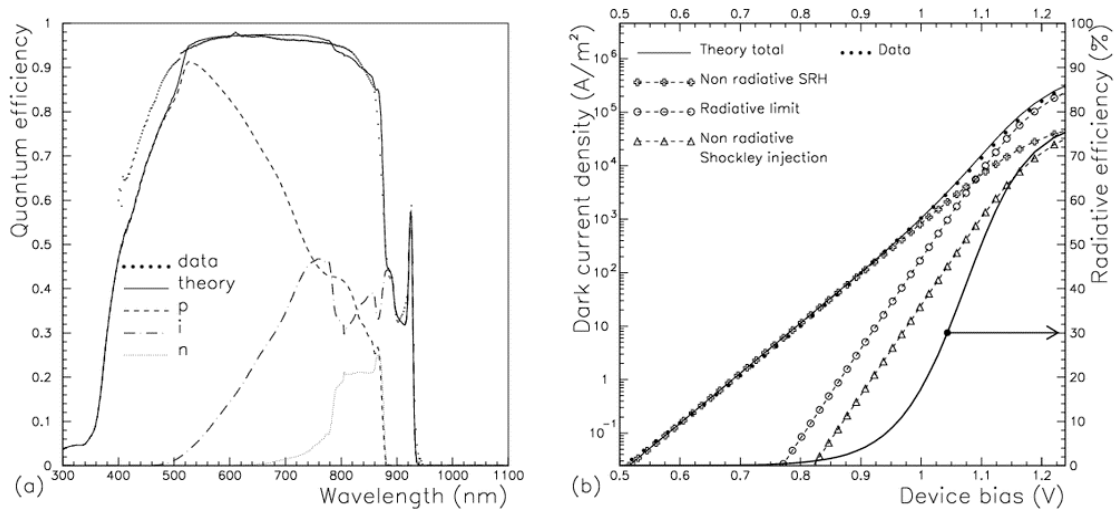

*Figure 5.3: (a) QE data and model for a QWSC PIN with GaAs barriers and GaInas well and (b) dark current fit together with radiative efficiency peaking at 75% at flat band and active area efficiency 22.5%*

## 3.1 Bulk and quantum well cells

The QE modelling and data for the bulk PIN is shown in figure 5.2a, showing a good fit to the data overall. The photocurrent contributions are broken down into *n* and *p* doped layer response, including the depleted sections of these layers, and the intrinsic or nominally undoped layer. The contributions show the expected dominance of the front-most p layer, particularly at short wavelengths, followed by the intrinsic n-type base layer, in order of decreasing light intensity as a function of depth.

The QE of the QWSC in figure 5.3a shows the same features as the PIN for emitter and base regions. The more intricate contribution from the wells merits some discussion. The wells are deeper than might be expected from the bulk bandgaps once strain shifts on the band-gap energy are taken into account, which, in this case shift the barriers upwards by about 50meV to 1.59eV or 782nm, and wells down by about 40meV to 1.26eV. Confinement increases the effective well gap to approximately 1.33eV or 926nm. The structure visible from the dominant bulk GaAs band-edge at 874nm to the strained barrier gap at 782nm is the higher quantum well bound states identified by the model as heavy hole states with just one light hole state lightly bound at the top of the well. The peaks visible in the intrinsic $i$ region response are therefore bound excitonic states of heavy hole and electronic states identified by the model but not reported in detail here.

The main conclusions we draw from the model is the illusory nature of the apparent close agreement between bulk and QWSC responses, since it masks the fine detail that may be revealed by explicit evaluation of the response of the different regions making up the solar cells.

The PIN recombination current is shown in figure 5.2b, from approximately half a typical Voc in order to see better the high bias behaviour. As described above, we see the non radiative Shockley-Read-Hall (SRH) marked by the empty crosses dominating at lower bias, with an modelled ideality of n=1.81 which is a striking match for the data, showing to what extent the common assumption of 2 is approximate.

The higher bias regime above 1V sees a change in slope which, although masked by the onset of series resistance, is clearly and very closely modelled by the explicitly non-radiative contribution to the Shockley injection current marked by the triangles, with an ideality of n=1. The final contribution is the upper limit or maximum possible radiative current from the cell is indicated by the empty circles. The assumption of upper limit of radiative efficiency does not lead to a visible overestimation of the dark current, despite the radiative current clearly approaching a few percent of the total dark current at high bias.

This last point is underlined by the explicit radiative efficiency (eq. 5.18) shown on the right hand axis. The bulk PIN cell is of sufficient quality to approach 20% radiative efficiency at high bias, which quantifies the extent to which this cell approaches the ideal limit. This important point shows that an ideality tending towards 1 cannot be taken to mean radiative dominance, as a consequence of the analysis we have presented which breaks down the Shockley injection current into radiative and non radiative components.

Returning to the QWSC in figure 5.3b, we observe the same trend from an SRH dominated regime with a slightly smaller ideality of n=1.78 again closely matching the data, to a regime of ideality n=1 looking remarkably similar to the PIN case. The fundamental difference however is clearly shown by the model: in the QWSC case, the contribution from the radiative recombination limit is slightly over an order of magnitude greater than the explicitly non radiative Shockley injection recombination current. The corresponding radiative efficiency for this case reaches 76% at high bias.

The important result of this analysis is that PIN bulk cells and QWSCs operate in different physical limits despite apparently similar dark current characteristics. The QWSC, radiatively dominated at biases above 1.05V, operates closer to the fundamental efficiency limit than the PIN cell, which this analysis shows is dominated by non radiative recombination in the ideality 1 regime. This must be qualified, however: as we have mentioned earlier, the structures are not optimal. As a result, it is a fortuitous consequence of material quality that the total dark current QWSC and bulk PIN devices are nearly equal, underlining the point made in the introduction concerning the greater importance of dominant carrier transport and recombination mechanisms rather than explicit performance issues

Having established this powerful notion of radiative dominance, we now focus on the obvious flip side, which is that the radiative dominance comes at the price of increased recombination. This draws attention back to the magnitude of the corresponding increase in photocurrent, which is not, in this non-optimal design, sufficient to counterbalance the increased recombination rate. As a result, the bulk PIN cell, despite it's less ideal, non-radiatively dominated recombination characteristic, is nevertheless the more efficient design simply because of the high absorption coefficient maintained to just a few nanometers in wavelength above the band-edge. In this specific case, however, the QWSC does perform more efficiently than the bulk cell due to secondary considerations, which are first the slightly better window and emitter design, and the better anti-reflection coating response for the QWSC.

This conclusion emphasises the introductory remarks on the non-ideal design of each cell and drawing attention to the physical regime the cells operate under rather than the conversion efficiency of each device. It furthermore echoes the observation made in the introduction regarding the relatively low well number, and resulting low well QE, with the first absorption continuum of just 33%, which, in turn, raises the issue of light trapping for increased absorption and consequences for the operating regime we shall cover next.

The strong conclusion however is that design specific issues apart, the QWSC is fundamentally a radiatively dominated design in a sense that the bulk PIN cannot be, unless non optimum, low doping levels are adopted in order to reduce the Shockley injection current. This demonstrated a promising high efficiency characteristic of this design.

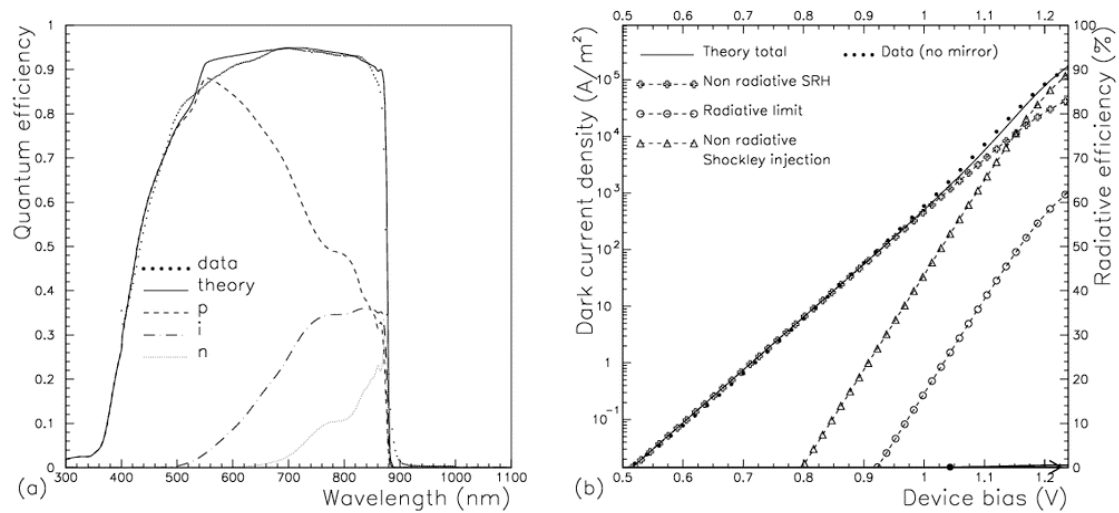

*Figure 5.4: Bulk PIN case QE theory (a) with back surface mirror compared with reference data without mirror, showing a negligible increase in QE near the band-edge. A similar dark current comparison (b) shows reduced radiative recombination, and hence a radiative efficiency peaking at just 1% at flat band. The mirror increases efficiency marginally from 21.0% to 21.3%*

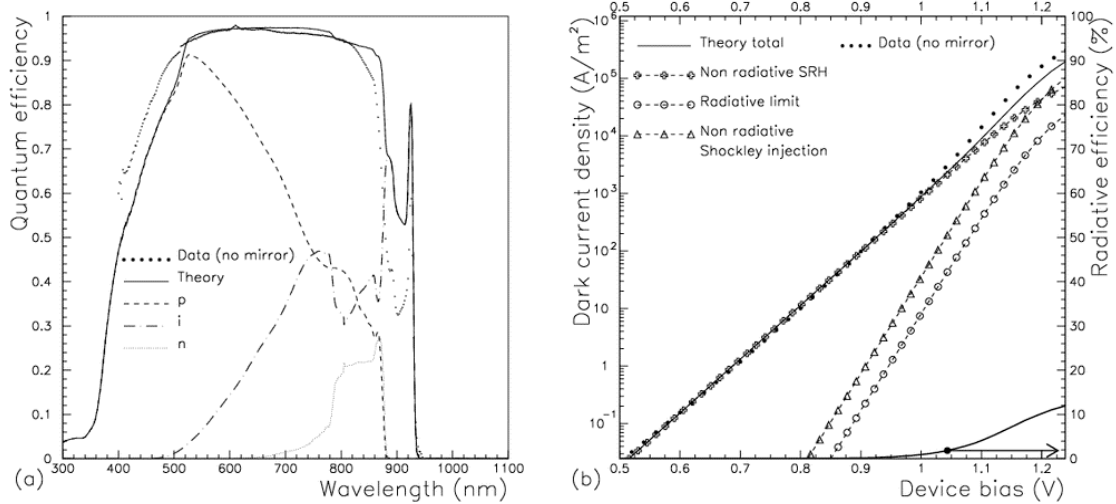

*Figure 5.5: SB-QWSC case (a) QE theory with back mirror compared with reference data without mirror showing significantly increased well response. The dark current theory (b) shows the mirrored SB-QWSC high bias dominated by the neutral layer non radiative Shockley injection current, and a radiative efficiency of the order of 12%. The SB-QWSC mirrored dark current is dominated by the higher bandgap charge-neutral. The mirror increases efficiency from 22.5% to 23.3%.*

## 3.2 Mirrors and restricted emission, or photon recycling

The previous section has emphasised the MQW low absorption as a weak point in the QWSC design. This problem is common to other thin film solar cell technologies however, and may be addressed by a range of geometric solutions such as textured surfaces and light management increasing the optical path length. Of these, the first and simplest solution is the back surface mirror as used in thin film solar cell photovoltaics, which is sufficient to illustrate the specific phenomenological features of the QWSC in the light of the analysis we have described. Coating surfaces with mirrors traps light and reduces light loss, but furthermore restricts light emission. The major advantage of this restricted emission, as we will see, in that coating a surface with a mirror cuts off the emission of light through that surface. As such, as ideal cell is one with the angle of light acceptance limited to the solid angle subtended by the spectral source, the sun, as described by de Vos (1992) and references therein, and complemented by Araujo & Martí (1994). This concept is closely related to photon recycling, differing by the increased absorptivity implied by photon recycling. In the ideal limit of an opaque cell and minimal spectral acceptance solid angle, the two concepts give the same radiatively limited efficiency.

In the case of a bulk PIN cells and QWSCs, the main difference is the transparence of the neutral regions. In the QWSC cell, the neutral layers are transparent to light emitted by the wells, whereas the bulk cell neutral layers of the bulk PIN are efficient absorbers of the *i* region emission. We will consider both QWSC and bulk PIN cells fabricated on substrates transparent to the luminescence in order to focus the description on the interaction between the active layers and the luminescence, rather than the less fundamental interaction with the substrate. We choose unit back reflectivity $R_b=1$ for similar reasons, but the results described apply to arbitrary non-zero back reflectivities.

Considering first the bulk PIN cell, figure 5.4a shows the modelled QE with a back mirror and, for reference, the same data as in figure 5.2a. Given the high absorption coefficient of

GaAs, the bulk cell absorbs nearly all incident light on the first pass. As a result, the QE with a back mirror shows only a slight increase in QE for wavelengths within a few nm of the band-edge, where the absorption coefficient is weakest. The dark current modelling in figure 5.4b however shows a marked difference to figure 5.2b, in that a strong reduction in radiative current is seen, whilst the non-radiative Shockley injection and Shockley-Read-Hall space-charge region currents are unchanged. The radiative efficiency therefore is reduced to slightly less than one percent as shown by the arrow on figure 5.2b. This is a consequence of the reduced net luminescence to the back surface, since that luminescence is reflected back into the cell by the back mirror according to equation 5.15. The second term in that expression relating to partial transmission through the back surface is reduced, while the third term relating to total transmission through the back surfaces vanishes for unit back reflectivity $R_b=1$. The net result is a reduction in the bulk PIN dark current, which is however relatively minor given the non-radiative dominated character of this cell. The practical advantage of the back surface mirror is, as might be expected for this opaque and non-radiatively dominated cell, relatively minor: The modelled efficiency increases marginally from 21% to 21.3% consistently with the minor impact on cell performance.

Figure 5.5a shows the corresponding effect on the QE of the QWSC. The quantum wells with the help of the mirror increase the short circuit current by 7% compared with a bulk control where wells and barriers are replaced by bulk GaAs, to 300.5A/m2 active area density under global spectrum AM1.5G. This furthermore represents a 3% increase over the QWSC case mentioned in the previous section without a mirror. It is clear, however, that the absolute level remains well short of the bulk QE, a level the quantum well must achieve in order to approach the radiative efficiency limits that it tends towards.

Concerning the radiative efficiency limit, figure 5.5b shows a striking reduction of dark current and a reversion to explicitly non-radiatively dominated Shockley injection dominance. The addition of the back surface mirror has the effect of reducing the luminescence by cutting off radiative losses towards the back of the cell. Neglecting emission in the small in-plane solid angle, the only remaining net loss path is though the small escape cone through the front of the solar cell, typically of some 17 degrees. The removal of the much greater 180 degree loss angle at the back surface translates as a reduction of approximately an order of magnitude in net radiative recombination that can occur. The balance of luminescence, which is still inevitable emitted via the Planck grey-body law, is re-absorbed, leading to re-emission, and so on, according to the formalism developed earlier in equation 5.15: This, in the dark, is equivalent to photon recycling, and is the mechanism whereby the back surface mirror transforms the QWSC from a radiatively dominated device to a non-radiatively dominated device.

Although at the first glance this appears a step backward, the comparison of figures 5.5b and 5.2b shows us that the QWSC with a back mirror is in fact dominated by the explicitly non radiative Shockley injection recombination rate in the charge neutral layers at sufficiently high bias, or equivalently, high current levels, such that SRH recombination no longer dominates. Such current levels are routinely achieved by cells operated under concentration of a few hundred suns as reviewed by Barnham et al. (2010). The consequence is that the QWSC can be operated in a regime where the fundamental absorption edge is determined by the space charge region, but the recombination mechanism remains determined by the higher bandgap charge neutral regions.
This important result is the second key advantage of the QWSC introduced earlier in this chapter, and suggested by Barnham and Duggan (1990). This analysis however pins down this conceptual disconnect between absorption edge and dominant recombination mechanism to the case where non-radiative recombination is present. More precisely, the ideal radiative case of pure, defect free material will show no such dominance of charge neutral recombination current, since this will have been reduced to a radiative level determined by it's gap. In this case, the cell will return to being dominated by the fundamentally greater radiative recombination current from the lower gap region. This limiting case, however, is clearly not one in which a working solar cell, which requires defects (dopants) in order to set up the built-in potential at the root of photovoltaic action.

In light of these points, it is clear that that the disconnect between observed dark current dominance by the higher gap bulk regions remains consistent with theoretical work that has suggested that the structure, while remaining subject to the generalised Planck law and corresponding efficiency limits, might offer routes to higher efficiency (de Vos (1992), and Araujo & Martí (1994)).

We conclude with a mention of experimental results reported by Johnson et al. (2007). This work has reported the observation of the phenomenon of reduced radiative recombination in SB-QWSCs with Bragg reflectors incorporated at the back of the cell, and demonstrate the dark current reductions of the order of those reported here.

## 4 Conclusions

Solar cell efficiency potential remains far greater than the best achieved the lab. The analytical model described in this chapter helps shed light on this by linking the solar cell performance in the light and in the dark, and using this to ascertain the dominant losses responsible for non-dealities. The set of approximations enabling this analytical methodology make it all the more important to minimise the free parameters. Consequently, the comparison of data and model crucial in determining what conclusions can justifiably be made. The model described satisfies both these conditions: the only free parameter is the Shockley-Read-Hall lifetime in the space-charge region, and furthermore, the modelled SRH ideality of slightly less than 2 agrees well with experiment.

The more interesting high bias regime of ideality one corresponding to higher efficiency concentrator current levels is however free of any fitting parameters. Close agreement between model and experiment is seen in both bulk and quantum well cases.

In this higher bias ideality one regime which is the focus of this chapter, some remarkable behaviour is revealed by comparing bulk and quantum well samples. The modelling and resulting analysis shows allows explicit estimates of the bias-dependent radiative efficiency of both classes of device.

It first transpires that the bulk device is non-radiatively dominated despite the transition from dark current ideality slightly below 2 to Shockley injection ideality 1. The radiative fraction in this high quality GaAs cell is only of the order of 20% at most, as the cell approaches flat band. The QWSC in the same regime is shown to be approximately 75% dominated by radiative recombination.

The analysis of samples with back surface mirrors reveals a quite different operational regime. The bulk PIN cell performance remains relatively unchanged, as expected for a nearly opaque structure which is dominated by non-radiative recombination. The QWSC however changes radically. First, the photocurrent increase is significant, which is a consequence of the low well absorption, and emphasising why light trapping is important for these structures. Secondly, and more significantly, the QWSC changes from a radiatively dominated to a non radiatively dominated regime. Examination of the contributions to the recombination current from the different regions of the cell shows that the net radiative recombination is suppressed by an order of magnitude through a combination of photon recycling and restriction of emission solid angle. As a result, the explicitly non radiative part of the Shockley injection current dominates the dark current. In other words, the dark current is primarily determined by the higher bandgap bulk charge neutral layers, rather than by the lower MQW layer.

The startling conclusion of this is that the Voc of the mirror backed QWSC is partly decoupled from the absorption edge. That is, the open circuit voltage of the mirror backed QWSC is determined by the high gap charge neutral layers, while the absorption edge is determined by the lower gap quantum well region. This is the original idea proposed by Barnham and Duggan (1990), albeit in a more restricted sense, in that this only holds true for doped, and therefore non radiatively dominated structures: in a structure where only radiative recombination is present, both PIN and QWSC designs revert to the Planck grey-body radiative limit, but without, however, the benefit of a doped junction to enable photovoltaic action and carrier collection.